\documentclass[twocolumn]{aastex63}
\usepackage{graphicx}
\usepackage{txfonts}
\usepackage{setspace} 
\usepackage[inline, shortlabels]{enumitem}

\def\ie{{\textit{i.e.,} }}
\def\be{\begin{equation}}
\def\bea{\begin{eqnarray}}
\def\eea{\end{eqnarray}}
\def\ee{\end{equation}}
\def\gmin{\Gamma_{\rm min}}

\def\Rph{R_{\rm ph}}

\def\tj{\theta_{\rm j}}

\def\tl{\theta_{l}}
\def\te{\theta_{\rm out}}

\def\g0{\Gamma_{0}}
\def\to{\theta_{\rm o}}
\def\Q{\cal{Q}}

\received{}
\revised{}
\accepted{}
\submitjournal{ApJL}

\shorttitle{Unified theory of spectral lags in GRB prompt phase}
\shortauthors{Vyas, Pe'er and Iyyani}
\graphicspath{{./}{figures/}}

\begin{document}
\setstretch{1.0}
\title{Unified theory of negative and positive spectral lags in GRB prompt phase due to shear Comptonization from a structured jet}

\correspondingauthor{Mukesh Kumar Vyas} 
\email{mukeshkvys@gmail.com}


\author{Mukesh Kumar Vyas}
\affiliation{Bar Ilan University, \\ Ramat Gan, Israel,
 5290002}
\author{Asaf Pe'er}
\affiliation{Bar Ilan University, \\ Ramat Gan, Israel,
 5290002}
 \author{Shabnam Iyyani}
\affiliation{School of Physics, IISER Thiruvananthapuram, Kerala, India, 695551}





\begin{abstract}
Positive spectral lags are commonly observed in gamma-ray burst (GRB) prompt phase where soft photons lag behind hard ones in their spectral studies. Opposite to this pattern, a fraction of GRBs show a negative spectral lag where hard photons arrive later compared to soft photons. Similarly, recent Fermi-LAT observations show a late onset of high-energy photons in most GRB observations. A fraction of GRBs show a transition from positive to negative lags. Such negative lags and the spectral lag transition have no convincing explanation. We show that a structured GRB jet with velocity shear naturally produces both positive and negative spectral lags. 
The high-energy photons gain energy from repeated scattering with shearing layers and subsequently escape from higher altitudes. Hence, these photons are delayed compared to soft photons producing a negative spectral lag. The inner jet has no shear and a positive lag appears providing a unified picture of spectral lags in GRBs. 
The theory predicts a flip in spectral lag from positive to negative within the evolution of the prompt phase. Comparison of the observed lags with the prediction of the theory limits the possible range of GRB jet Lorentz factors to be a few tens. 
\end{abstract}
\keywords{High energy astrophysics; Gamma Ray Bursts; Theoretical models}

\section{Introduction}
\label{sec_intro}

From the first evidence by \cite{1995A&A...300..746C}, a delay between the soft and hard photons (in the range of few keV - few hundred keV) known as spectral lag is extensively observed in the prompt phase of gamma-ray bursts (GRBs). The spectral lags in GRBs are inferred either by light curve fitting method \citep{2008ApJ...677L..81H} or cross-correlation function (CCF) \citep{1997ApJ...486..928B}. CCF is primarily used where the light curves are compared between different energy bands \citep{2000ApJ...534..248N, 2004ApJ...606..583L, 2012PASP..124..297L, 2005ApJ...619..983C, 2010ApJ...711.1073U}.

Studying around 2000 GRBs from the BATSE catalogue, \cite{2007ApJS..169...62H} found that around 70\% of the long GRBs (lasting for tens of seconds) have a positive delay where the photons evolve from hard to soft energies. 
15\% of the bursts had a negative spectral lag where soft photons precede the hard photons.
\cite{2021ApJ...922...34C} found a positive lag in  89.8\% GRBs while 9.5\% bursts had a negative spectral lag. 
In addition, several reports of GRBs show both positive and negative lags. For example, \cite{2014ApJ...782..105R} showed that GRB 060814 exhibits both positive and negative spectral delays. Similar spectral transition from positive to negative lag is reported in other cases such as GRB 160625B \citep{2017ApJ...834L..13W, 2022PASA...39....1G}, GRB 160625A \citep{2023ApJ...942...67L} and 190530A \citep{2022MNRAS.511.1694G}. \cite{2022ApJ...935...79L} reported 32 GRBs that exhibit a transition from positive to negative spectral delay.
In these studies, the obtained lags range from milliseconds up to a few seconds.  Moreover, recent results from Fermi-LAT data reveal that negative spectral lags generally persist in GRBs at even longer times, extending up to several minutes where the high energy photons, in the MeV-GeV energy range, show a late onset \citep{2013ApJS..209...11A, 2019ApJ...878...52A, 2019ICRC...36..555B}.

Most short GRBs are known to have no spectral lag \citep{2000ApJ...534..248N, 2010ApJ...711.1073U, 2006ApJ...643..266N} while a negative lag is reported in about 17\% of the short bursts \citep{2006MNRAS.367.1751Y}. This makes spectral delay a parameter distinguishing between short and long GRBs. However, as these are short-duration bursts lasting for less than 2 seconds, the delays between various energy channels are debated and considered too short to be reliable \citep{2007ApJS..169...62H}.

The exact origin of the spectral lags is not fully understood. The positive lag was attributed to curvature effect \citep{2002ApJ...578..290R, 2004ApJ...614..284D, 2014Sci...343...51P, 2016ApJ...825...97U} or, alternatively, to spectral evolution during the burst \citep{2003ApJ...594..385K, 2005ApJ...625L..95R}. Recently, a positive spectral lag was demonstrated to naturally arise from the curvature of a backscattering photosphere of a propagating expanding GRB jet \citep{2021ApJ...908....9V}. 
A physical understanding of negative spectral lags is a bigger challenge for theoretical modelling \citep{2018ApJ...869..100U, 2021MNRAS.505.4086G}. It was suggested by \cite{2018JHEAp..18...15C} that this lag may result from Compton scattering of the low energy emitted photons with an external medium such as a cloud surrounding the GRB jet.

In this work, we show that both positive and negative lags are a natural outcome of a structured jet. 
In a previous work, we established that photon propagation in a sub-photospheric region of a structured jet naturally results in a power law-shaped spectrum at high energies, originating from a Fermi-like energy gain process \citep{2023ApJ...943L...3V}. The photons gain energy through repeated scattering within the shear layers of the relativistic jet. 
Here, we show that such a model naturally produces a negative spectral lag when applied to the time evolution of the GRB prompt phase. The photons that gain higher energy scatter multiple times within the shear layers and hence lag behind the soft photons which escape earlier to produce a negative spectral lag. We provide here both an analytical calculation as well as an independent study based on Monte Carlo simulations of photon propagation within the structured GRB jet.

The predicted temporal evolution of high-energy photons is consistent with the observations, thereby explaining the negative spectral lag as well as the transition between the positive and the negative lags. 
In section \ref{sec_model}, we explain the model and the physical assumptions we use. We describe results in section \ref{sec_results} and conclude the work with its significance in section \ref{sec_conclusions}.

\section{Model and physical mechanism}
\label{sec_model}

\subsection{Structured jet}
\label{sec_jet_stricture}
Based on the simulations of an erupting jet from a stellar collapse \citep{2003ApJ...586..356Z, 2021MNRAS.500.3511G, 2024MNRAS.531.1704G}, we consider a structured jet with a (steady) angle-dependent Lorentz factor of the form \citep[see][]{2013MNRAS.428.2430L, 2017IJMPD..2630018P},
\be 
\Gamma(\theta)= \gmin+\frac{\Gamma_{0}-\gmin}{\sqrt{\left(\frac{\theta}{\theta_{\rm j}}\right)^{2p}+1}}.
\label{eq_gamma_1}
\ee 
Here $\theta$ is the polar angle in a spherical coordinate system ($r,\theta, \phi$) with the origin being at the centre of the burst. The system possesses azimuthal symmetry. The jet has a constant and maximum Lorentz factor near the jet axis for angles $\theta \ll \theta_j$ and asymptotically reaches $\gmin$ at large angles, $\theta \gg \te = \tj \Gamma_0^{1/p}$. We denote the outer boundary of the jet by $\te$, assuming no matter ejected at $\theta>\te$. Hence the photons that happen to reach angles beyond $\te$ are assumed to reach the jet boundary and escape. We stress that this is only a tiny fraction of the escaping photons, who, by large, escape at the photosphere (see below).
The isotropic equivalent luminosity $L$ of the jet is assumed to be independent of the jet angle \ie $dL/d\Omega = L/4\pi$. The temperature at the jet base $r_i$ is
\be 
T_0=\left[\frac{L}{4 \pi r_i^2 a_r c }\right]^{1/4}.
\label{eq_T_0}
\ee 
Here $a_r$ is radiation constant and $c$ is the speed of light. Followed by adiabatic cooling due to jet expansion, the comoving temperature of the jet first drops as $r^{-1}$ until the saturation radius, $r_S \simeq r_i \Gamma$, and varies as $T'=T_0(r_i/r_s)(r_s/r)^{2/3} = T_0 (r_i/\sqrt{\Gamma}r)^{2/3}$ at larger radii \citep{2012MNRAS.420..468P, 2013MNRAS.428.2430L}.

While the jet base is often taken to be the size of the newly-formed black hole, i.e., $r_i \approx 10^6$~cm, we point out that this is most likely not the case for long GRBs. As the GRB jet drills its way through the collapsing star (presumably, a Wolf-Rayet star), it encounters several recollimation shocks that slow it \citep{2000ApJ...531L.119A, 2009ApJ...700L..47L, 2011ApJ...740..100B, 2013ApJ...777..162M,2014MNRAS.442.2202L, 2015JHEAp...7...17L}. Eventually, the jet is brought to a near halt before it re-accelerates \citep{2022Univ....8..294V}. For typical GRB parameters, this occurs at radii in the range $10^9 - 10^{10}$~cm \citep{2007ApJ...666.1012T, 2013MNRAS.433.2739I, 2015ApJ...813..127P}. Thus, the effective jet base is at this radius. We therefore consider $r_i = 10^{10}$~cm in this work.


\subsection{Numerical simulation}

Our numerical model is based on the Monte Carlo method.  The jet is assumed to propagate in the radial direction with an angle-dependent Lorentz factor given by Equation \ref{eq_gamma_1}. Because of the conical expansion, the density of particles drops with radius, and photons emitted deep inside the flow will eventually escape the expanding jet, once they reach the photosphere.  

Photons are injected deep inside the jet (i.e., in regions of high optical depth) with random directions in the jet comoving frame. The photons go through repeated scattering with the electrons in the jet. The code traces the propagation of individual photons between two scatterings, as well as randomly generates a scattering event. In each scattering, the photon four-vector is transformed from the comoving fluid frame to the local electron's rest frame where an individual interaction is calculated and back; finally, it is transferred to the observer's frame. 

After scattering with an individual electron, the photon travels a distance $\delta l$ along the scattered direction before the next scattering. The probability for the photon to have a consecutive scattering after a distance $\delta l$ is $\exp^{-\tau}$, where $\tau$ is the optical depth for scattering by the local fluid element, calculated along the photon propagation direction. This optical depth is randomly drawn from a uniform distribution. Furthermore, the optical depth for escape, $\tau_{\rm esc}$ is calculated at each time step. If $\tau>\tau_{\rm esc}$, the photon escapes the jet. Otherwise, the distance $\delta l$ is calculated such that the local optical depth between two consecutive scatterings is equal to $\tau$. The same process continues until the photon escapes the system.  The code traces the time it takes a photon from injection to escape.


All the photons are injected into the plasma at radius  $r = \Rph/20$, where $\Rph = \Rph(\theta)$ is the photospheric radius (See appendix A). We find that the photon injection radius is deep enough to capture all the essential signals, and deeper injection does not affect our results, but only increases the computational time.
We find that nearly all photons escape the plasma from within the jet, i.e., $\theta < \te$ at the photon escape location.
Photons that leave the system early are the ones that do not go through a large number of scattering inside the shearing plasma but rather scatter within the jet core ($\theta < \tj$). On the other hand, photons that go through repeated scattering with the shearing layers ($\theta > \tj$) are accelerated to high energies and escape from higher latitudes. Hence they are subsequently delayed.
For a detailed description of the model, 
see \cite{2008ApJ...682..463P, 2013MNRAS.428.2430L} and \cite{2023ApJ...943L...3V}.

\subsection{Mechanism of photon energy gain and subsequent delay}

For a photon undergoing repeated scattering through electrons in a cold jet, we provided an analytical expression for the average energy gain per scattering $g(r,\theta)$ in \citet{2023ApJ...943L...3V}. This gain is defined as the ratio of photon energies after and before a scattering event as observed in the lab frame,
\be 
g(r,\theta) \approx\frac{1}{2}\left\{1+\left[1+ \frac{\partial \log \Gamma}{\partial \theta }\delta \theta\right]^2 \frac{1}{\left(1+a\right)^2}\right\}.
\label{eq_grtheta}
\ee
Here, $a=\lambda/r$ is the expansion parameter that considers energy loss due to the jet's adiabatic expansion. Here $\lambda$ is the local mean free path,
\be 
\lambda = \frac{\Gamma}{n_e'\sigma_T}.
\label{eq_lambda}
\ee 
The comoving electron density inside the jet is $n_{e}'=L/4\pi m_{p} c^3 \beta \Gamma^2 r^2$. The proton mass is $m_p$, and $\beta = \beta(\theta)$ is the local jet velocity normalized to the speed of light $c$. The angular displacement between two scattering is denoted by $\delta \theta$. Its dependence on the scattering location is approximated as \citep{2023ApJ...943L...3V},
\be 
\delta \theta = -\frac{a}{\Gamma (1+a)}.
\ee
For a jet profile with $p>2$, in the region $\te>\theta>\tj$ the energy gain overcomes the adiabatic energy losses and $g > 1$. 


With the help of this expression, and noting that in the shearing region, $\Gamma \sim \Gamma_0 (\theta/\tj)^{-p}$ (with $\gmin \ll \Gamma_0$), one can express Equation \ref{eq_grtheta} as
\be 
g(r,\theta) \approx \frac{1}{2}\left[1+\frac{[1+a(1+k)]^2}{(1+a)^4}\right].
\label{eq:5}
\ee 
Here, $a=a(r) \propto r$ is defined above, and  
$$
k= k(\theta) = \frac{p}{\Gamma_0 \tj}\left(\frac{\theta}{\tj}\right)^{p-1}
$$
is a function of the angle within the jet.

For plausible jet parameters (e.g., $\Gamma_0 = 100$, $\theta_j = 0.01 - 0.1$), in the shearing region, $\te>\theta>\tj$, $k$ can be as high as a few (even few tens), while close to the photosphere, the mean free path becomes comparable to the photospheric radius and $a\lesssim 1$. Using Equation \ref{eq:5}, one therefore finds that the the gain increases with radius, and obtains its maximal value close to the photosphere.




\subsubsection{Late arrival of energetic photons from the shearing region $\theta_j<\theta<\te$ and a negative spectral lag}
 
The photospheric radius, $\Rph$ depends on the polar coordinates $\{ \theta, \phi \}$ as well as the observer's angular location $\to$, $\Rph = \Rph (\theta,\phi;\to)$. For an observer located on the jet axis ($\to=0$), this radius is independent of the azimuthal angle $\phi$ and obtains its minimum in the inner jet region ($\theta < \tj$). It increases with angle at $\theta > \tj$ \citep[see figure 4 in][]{2013MNRAS.428.2430L}. For an off-axis observer, the angular dependence is more complicated and is calculated in Appendix A. 


While the photon's propagation direction is random in the comoving frame, the photons are coupled to the electrons below the photosphere and therefore are advected with the flow, gaining energy.
As we showed above, the gain increases with the radial coordinate, obtaining its maximal value at the last scattering. As a crude approximation which is nonetheless useful for physical insight, one may consider the photons to gain all their energy in a single scattering occurring at a single radius, namely the photosphere. The average gain at the photospheric radius $g(r,\theta)=g_a(\Rph,\theta)$ (the subscript `a' stands for average) is estimated using $r=\Rph$ in Equation \ref{eq_grtheta} [Also see the detailed description below Equation 10 in \cite{2023ApJ...943L...3V}].

\begin {figure}
\begin{center}
  \includegraphics[width=9cm, angle=0]{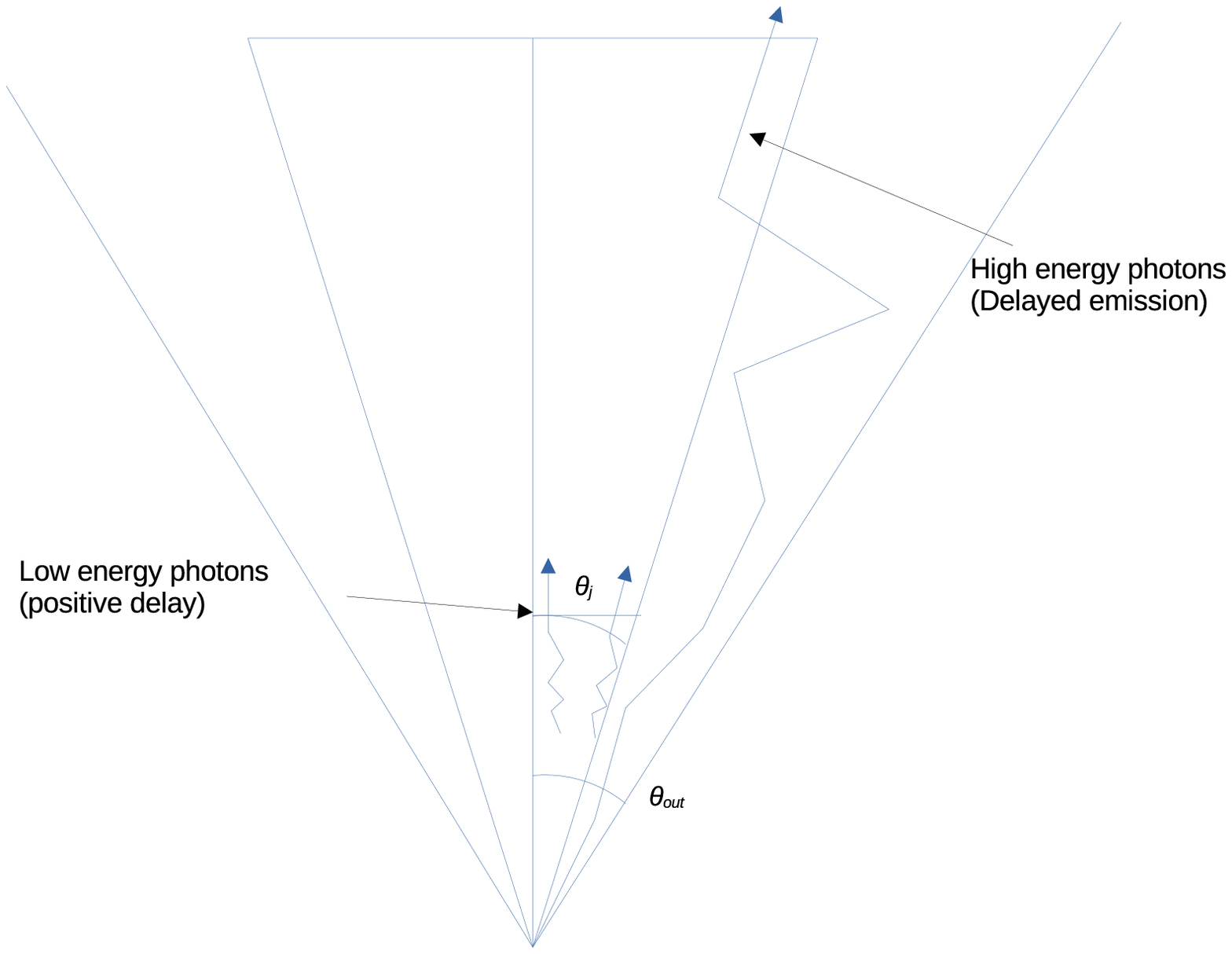}
   \includegraphics[width=9cm, angle=0]{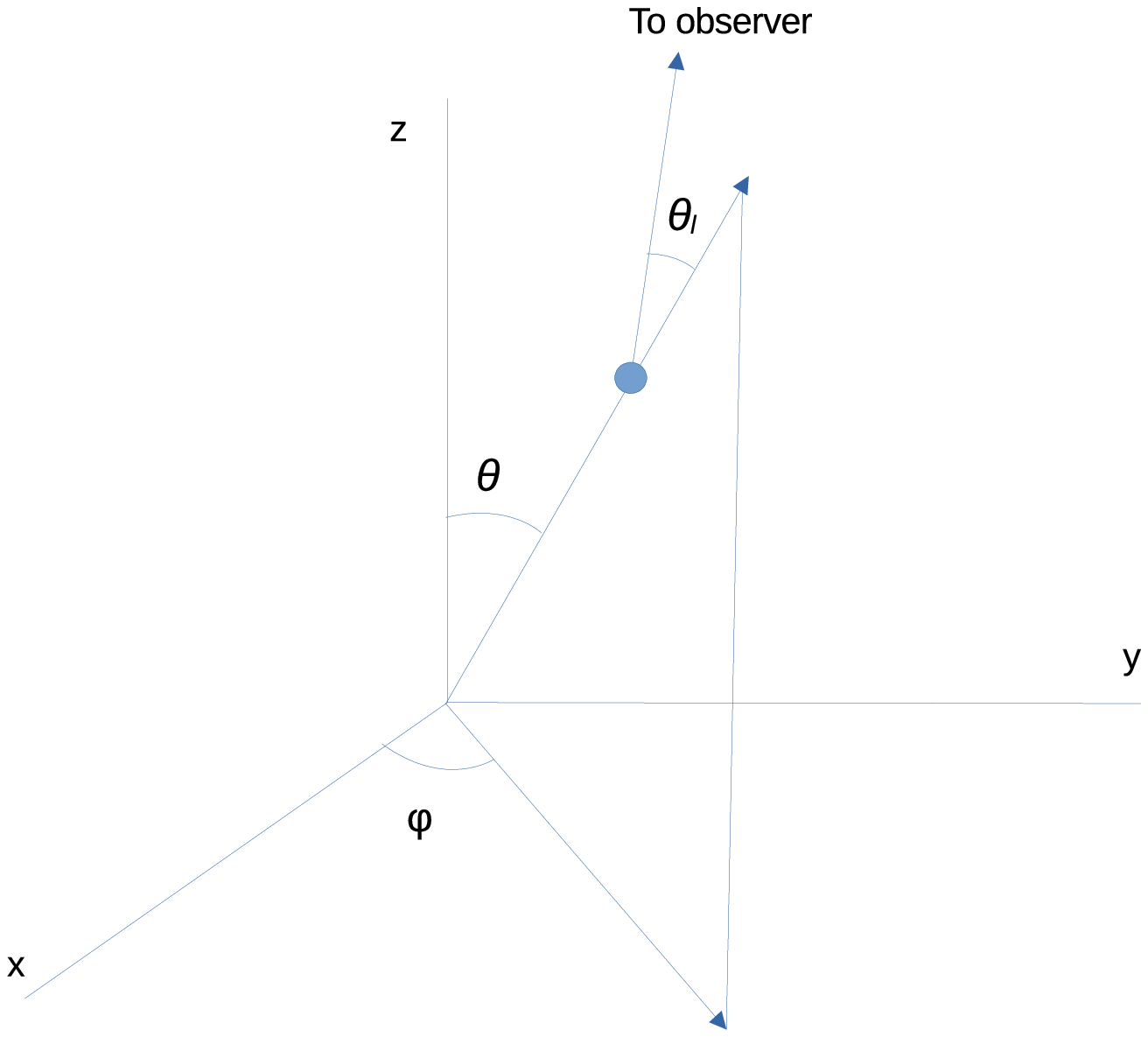}
\caption{Upper panel: Geometrical representation of the spectral lags. Positive lag is produced within $\te$ due to the curvature effect while negative lag is produced due to the late emission of high energy photons from the region bounded within $\tj-\te$,
 where the energy gain takes place. However, the high energy photons typically escape from the inner jet region ($\theta<\tj$) at the last scattering.
Lower panel: Geometry of the last scattering angle. $\tl$ is the angle between the radially moving electron and the observer's location. If the observer is assumed at $\to=0$ then $\tl=\theta$.
}
\label{lab_geom1}
 \end{center}
\end{figure}

As shown in \cite{2023ApJ...943L...3V},  while energetic photons propagate within the shearing region ($\theta>\tj$) and subsequently gain energy, most high energy photons escape through the inner jet region, $\theta< \tj$. This is due to the fact that the photospheric radius is concave, namely for $\theta_1 > \theta_2$, $\Rph (\theta_1) > \Rph(\theta_2)$ \citep{2008ApJ...682..463P, 2013MNRAS.428.2430L} (both angles are measured from the jet axis). 
Therefore, photons that propagate at $\theta>\tj$ and (randomly)  scatter into the inner jet region consequently escape. This scenario is demonstrated in the top panel of Figure \ref{lab_geom1}.
Denoting the photon (comoving) energy before scattering by $E'_i$, the observed energy $E$ after the scattering is therefore 
\be 
E=\frac{g_a(\Rph,\theta) E'_i}{\Gamma_0(1-\beta_0 \cos \tl)}.
\label{eq_E_escape}
\ee 
Here, the gain $g_a$ is calculated at the photon propagation location just before its last scattering, $R_{ph}(\theta)$, while the Doppler boost is calculated from the inner jet region (i.e., $\beta=\beta_0$ is jet velocity normalized to the speed of light in the inner jet region, $\Gamma_0$ is the corresponding Lorentz factor) and $\tl$ is the angle between the radial direction of the electron and the observer (see bottom panel of Figure \ref{lab_geom1}). 

Due to the multiple scatterings, these photons are delayed with respect to a hypothetical photon emitted at the centre ($r=0$) and propagates towards the observer. To calculate this delay, 
we assume that the plasma velocity ($\beta$) is independent of the jet propagation radius (i.e., the Lorentz factor depends only on $\theta$, but not on $r$).
For those photons that propagate with the plasma at angle $\theta$ from the jet axis and last scatter towards the observer at angle $
\theta_l$, 
this delay is given by (see Figure \ref{lab_geom1}, bottom panel),
\be 
\Delta t^{ob.} =\frac{\Rph(\theta, \phi; \theta_o)}{\beta_0 c}\left[1-\beta_0 \cos \tl\right].
\label{eq_time}
\ee 
For a fluid element propagating along the jet axis and emitting a photon in the observer's direction, one finds $\tl=\to$. Similarly, for the specific case of an observer along the jet axis, $\tl=\theta$. In general, one finds  $\theta_l = \left( \to^2 - 2 \to \theta \cos \phi + \theta^2 \right)^{1/2}$.

In the specific case of an on-axis observer ($\theta_o=0$), the photosphere possesses azimuthal symmetry and $\tl=\theta$. In this case, the time difference is
\be 
\Delta t^{ob.}=\frac{\Rph(\theta)}{\beta_0 c}\left[1-\beta_0\cos \theta\right].
\label{eq_time_0}
\ee 
We estimate the photospheric radius $\Rph(\theta, \phi; \theta_o)$ as described in \cite{2024ApJ...972...40V} and in appendix A for a given observer's location $\to$, and calculate $g_a[r=\Rph(\theta; \theta_o, \phi)]$ using Equation \ref{eq_grtheta}. Equations \ref{eq_E_escape} and \ref{eq_time} allow us to estimate the observed photon energy $E$ as a function of observing time $t = \Delta t^{ob.} - \Delta t^{ob.}_0$. Here  $\Delta t^{ob.}_0$ is the observed time of the first photon from the photosphere at its minimum $\Rph=\Rph{_0}$. For an on-axis observer, it is,
\be 
\Delta t^{ob.}_0 = \frac{\Rph{_0}}{\beta_0 c}\left[1-\beta_0\right].
\label{eq_dt0}
\ee 

\subsubsection{Positive spectral lag by low energy photons from inner jet region $\theta<\theta_j$}
In the inner jet region ($\theta<\theta_j$), the Lorentz factor is angle-independent and has no shear, implying that the scattered photons do not gain energy. The photons do lose energy due to the adiabatic expansion as they propagate along the expanding jet.

Consider first an observer along the jet axis. The first photon received by the observer escapes close to the jet axis, $\theta \approx 0$. Photons propagating at larger angles escape at later times, as given by equation \ref{eq_time_0}. 

The energy of these photons is estimated identically to the calculation that led to Equation \ref{eq_E_escape}, the only difference being that photons propagating at $\theta < \theta_j$ do not gain but only lose energy, therefore $g_a < 1$ (see Equation \ref{eq:5} with $k \ll 1$) and one can write 
\be 
E=\frac{g_a E_1'}{\Gamma_0(1-\beta_0\cos(\theta_l))} = \frac{g_{a0}E_i'\Rph{_0}}{\Gamma_0\beta_0 c \Delta t^{ob.}} {\rm ~~~~~(for~\theta_o = 0)}
\label{eq_lag_positive}
\ee 
where the last equality holds for an on-axis observer, and we used Equations \ref{eq_E_escape} and \ref{eq_time_0}.
Here $g_{a0} = g_a(\Rph,\theta=0)$ is the energy gain at the jet axis $\theta=0$. 

The result of Equation \ref{eq_lag_positive} shows that later photons arrive to the observer with lower energies, as $E \propto {\Delta t^{ob.}}^{-1}$. This result can easily be understood, as later photons originate from higher angles, hence with a lower Doppler boost relative to an on-axis photon. All these photons are less energetic than the initially observed photon. This is also known as ``the curvature effect", leading to an observed positive spectral lag. 

A similar calculation for photons originating at angles $\theta > \theta_j$ (Equations  \ref{eq_E_escape} and \ref{eq_time_0}) reveals a more complicated result, as photons that originate from larger angles and hence are observed at a longer delay, on the one hand, are observed with lower Doppler boost hence with lower energies, but on the other hand they gain more energy due to the longer time they spend in the shearing region (the parameter $g_a$ in Equation \ref{eq_E_escape}). Overall, the energy gain dominates, and these photons are observed at higher energies than the initially observed photons. These energetic photons arrive later than the lower energy photons. Thus, overall, the low energy photons arrive early, causing the observed {\it positive delay}, while the energetic photons arrive later, resulting in the observed {\it negative delay}.

At time $\Delta t^{ob.}=\Delta t^{ob.}_0$, where the first photon reaches the observer, its observed energy defines the transition energy $E_{\rm Tr}$ that separates positive lags and negative lags. This energy is obtained using Equation \ref{eq_lag_positive} for $\Delta t^{ob.}=\Delta t^{ob.}_0$ with the help of Equation \ref{eq_dt0},
\be 
E_{\rm Tr} = \frac{g_{a0}E_i'}{\Gamma_0(1-\beta_0)} \approx 2 \Gamma_0 g_{a0}E_i' {~~\rm (for~~}\Gamma_0>>1).
\label{eq_Etr}
\ee 
As explained above, the energetic photons having observed energy $E > E_{\rm Tr}$ show negative lags due to their escape from higher altitudes.
Lower energy photons, having $E<E_{\rm Tr}$ are observed at a positive lag due to the curvature effect. The mechanism of positive and negative delays is represented in the top panel of Figure \ref{lab_geom1}.



\begin {figure}[h]
\begin{center}
 \includegraphics[width=7cm, angle=0]{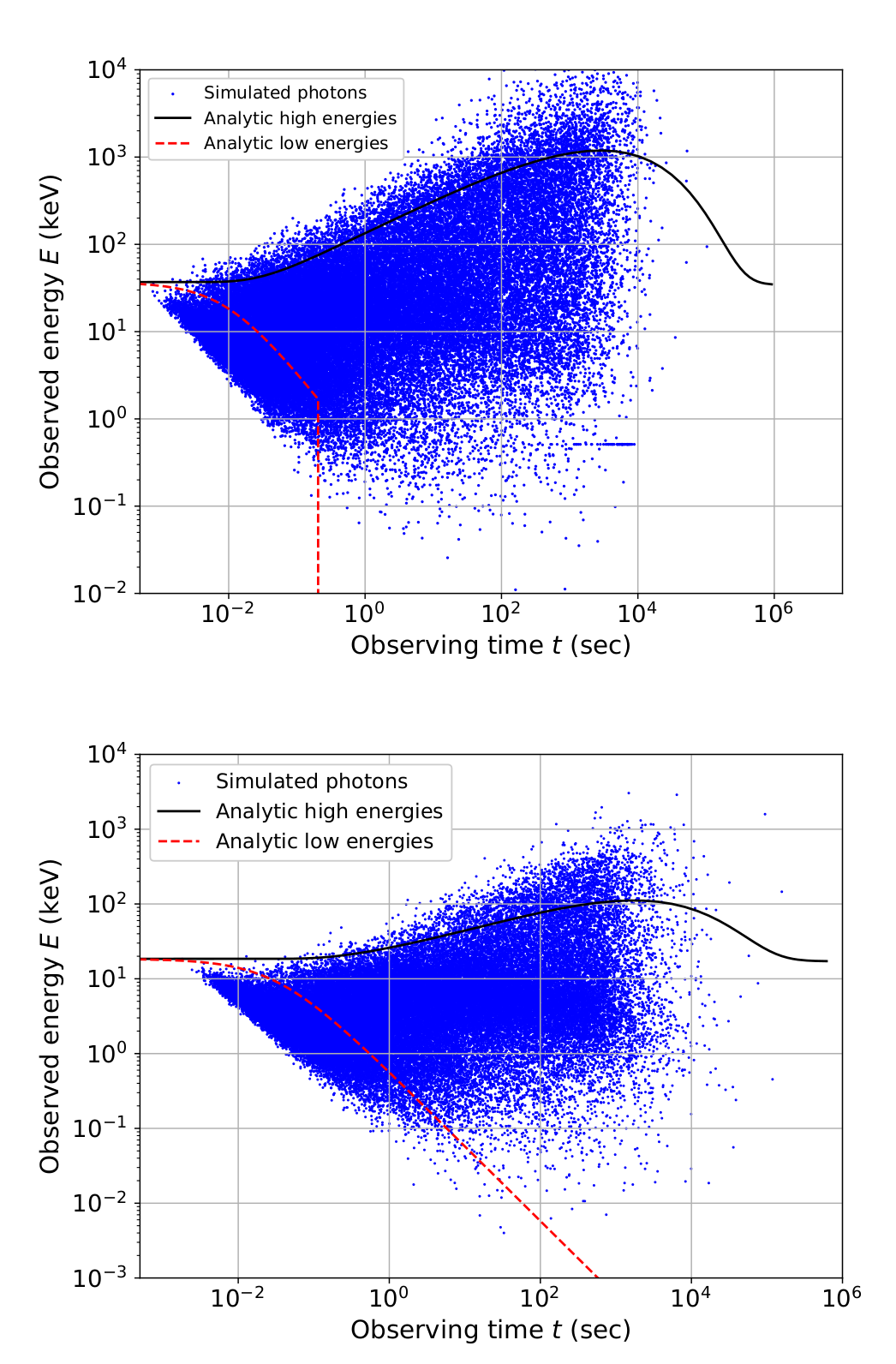}
\caption{Distribution of the observed photon energies $E$ over observing time $t$ for the observer situated along the jet axis for parameters $L=10^{52}$ ~erg~s$^{-1}$, $\tj=0.01$ rad, $p=4$ and $\Gamma_0 = 100$ (top panel) and $L=10^{51}$~erg~s$^{-1}$ $\tj=0.03$ rad and $\Gamma_0 = 50$ (bottom panel). The photons are injected with energies $E_i=10^{-3}m_ec^2$.
}
\label{lab_Photon_counts_g0_100_tj_0.01_L_1e52}
 \end{center}
\end{figure}
\section{Numerical results}
\label{sec_results}
To validate the analytical calculations presented above, we carried a Monte-Carlo simulation of photon propagation below the photosphere \cite[see details of the model in][]{2024ApJ...972...40V}.
We consider a cold structured jet with an angle-dependent Lorentz factor given by Equation \ref{eq_gamma_1} with $\gmin=1.2$ throughout this letter. Monoenergetic photons are injected at $r_i=\Rph/20$ evenly distributed at $\theta < \theta_{\rm out}$ with energies $E_i'=10^{-3}$ normalized to the electron's rest mass energy.
The photons propagate with the flow, and those propagating within the shear layers gain energy through repeated scattering within the shear (the region $\tj <\theta< \te$). 
These photons traverse a longer path before reaching the observer and hence are delayed.

In Figure \ref{lab_Photon_counts_g0_100_tj_0.01_L_1e52} we show the numerical results for $\sim 10^7$ photons.
Each dot represents the observed energy vs. observed time of an individual photon. The results are shown for two sets of parameters representatives of GRBs: (a) $L=10^{52}$~erg~s$^{-1}$, $\tj=0.01$~rad and $\Gamma_0 = 100$ (top panel) and (b) $L=10^{51}$~erg~s$^{-1}$, $\tj=0.03$~rad and $\Gamma_0 = 50$ (bottom panel). The observer is assumed to be located on the jet axis ($\to=0$) and the shear parameter is $p=4$.

The soft photons that escape from the photospheric radius within the inner jet ($\theta<\tj$) escape earlier and produce a positive spectral lag due to the curvature effect as explained above. On the other hand, high energy photons (having energies above $E_{\rm Tr}$) are delayed due to the lag induced by their propagation inside the shear layers. This can be seen as the more energetic photons appear later and with a more gradual distribution increase than the lower energy ones.


Analytically estimated energy evolution of photons is plotted by the black solid line (negative lag from Equations \ref{eq_E_escape} and \ref{eq_time}) and the red dashed line (positive lag following Equation \ref{eq_lag_positive}). At $t = 0$, both curves coincide at $E=E_{\rm Tr}$ (Equation \ref{eq_Etr}). Thus, the analytic approximations are validated by the numerical simulations.

\begin {figure}
\begin{center}
 \includegraphics[width=9cm, angle=0]{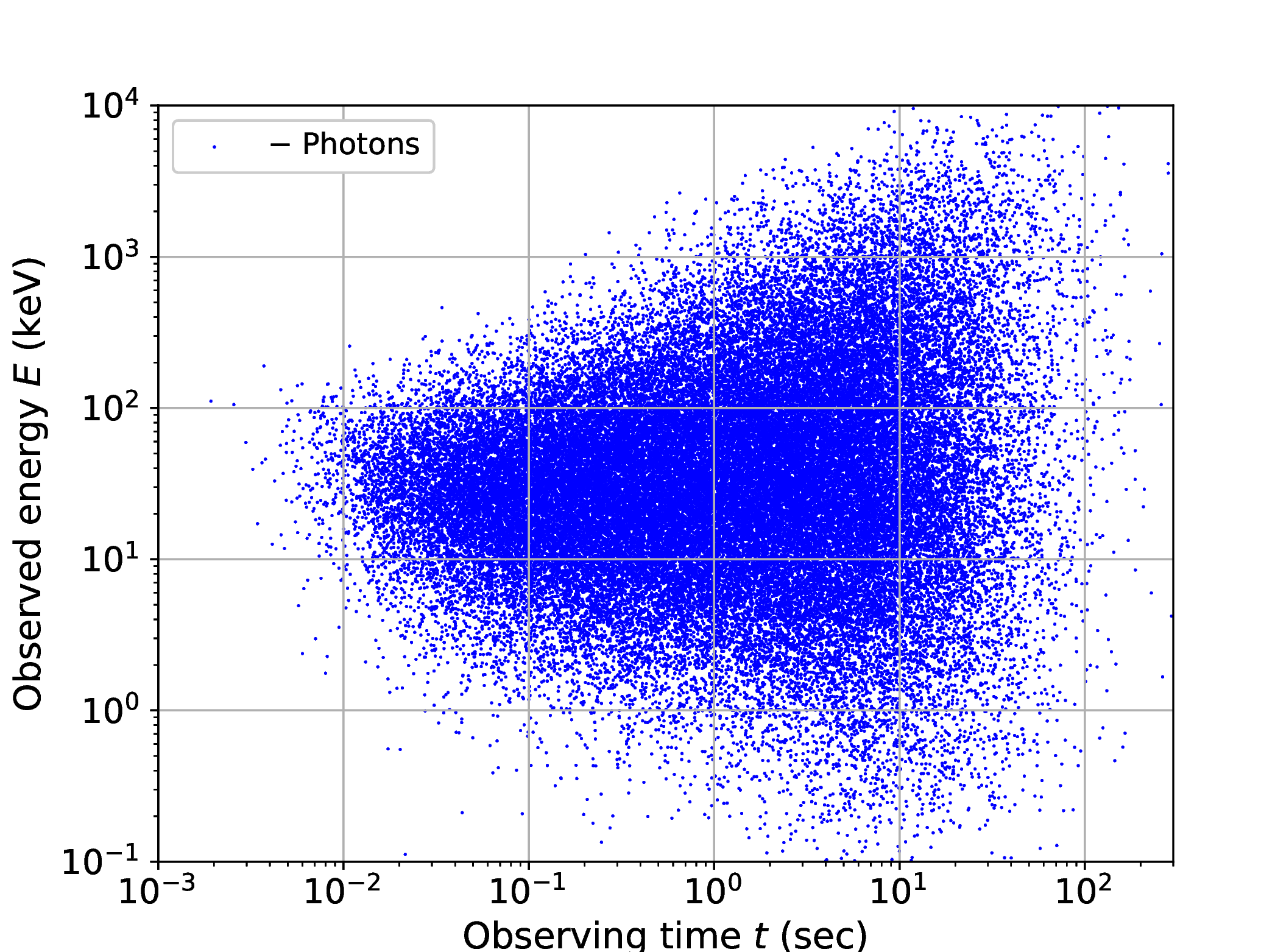}
\caption{Photon map in $E-t$ plane for $L=10^{50}$ ~erg~s$^{-1}$, $\tj=0.01$ rad, $\Gamma_0=30$, $p=4$, $\gmin=1.2$. The photons are launched at $r_{\rm inj}=\Rph/20$ following black body distribution with corresponding fluid temperature $T'(r_i, \theta)$.  
}
\label{lab_Photon_counts_g0_30_BB}
 \end{center}
\end{figure}

\begin {figure}
\begin{center}
 \includegraphics[width=9cm, angle=0]{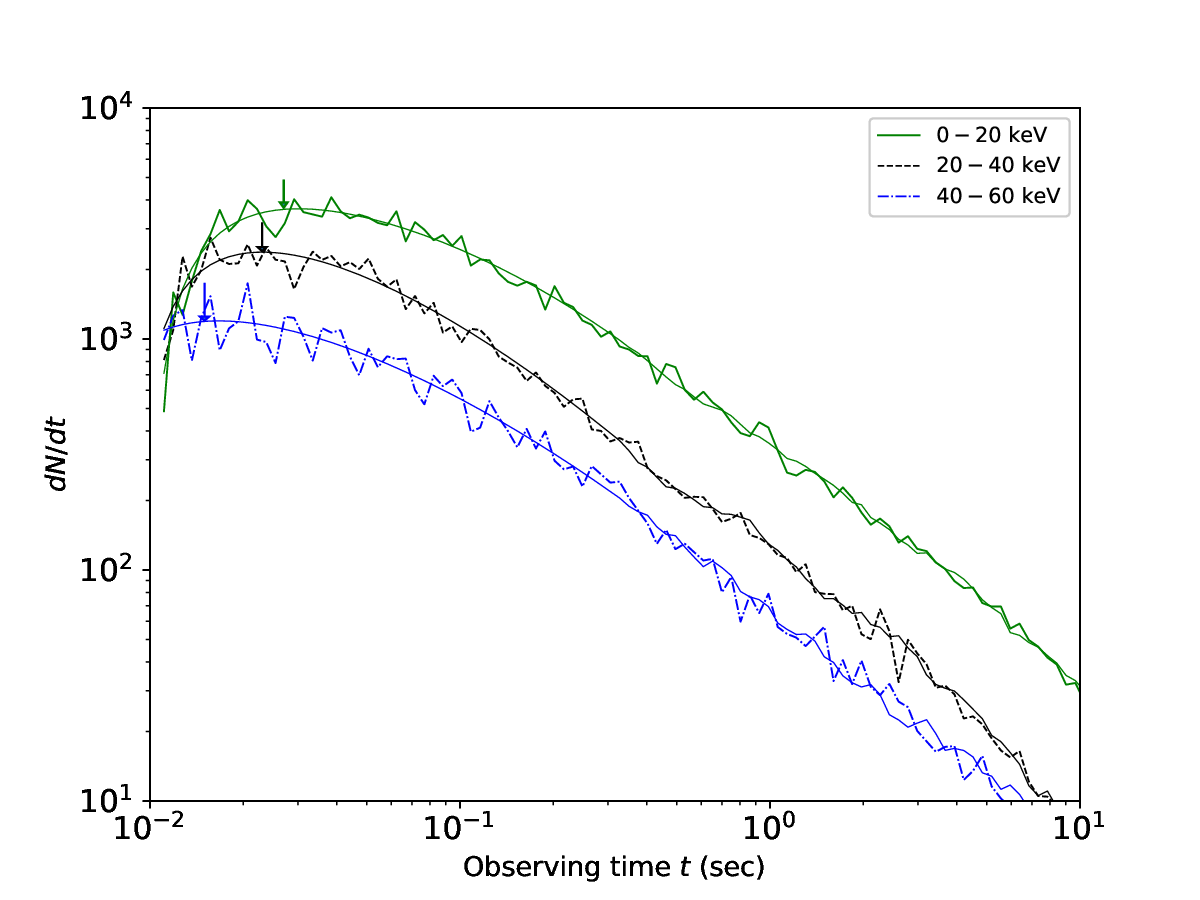}
  \includegraphics[width=9cm, angle=0]{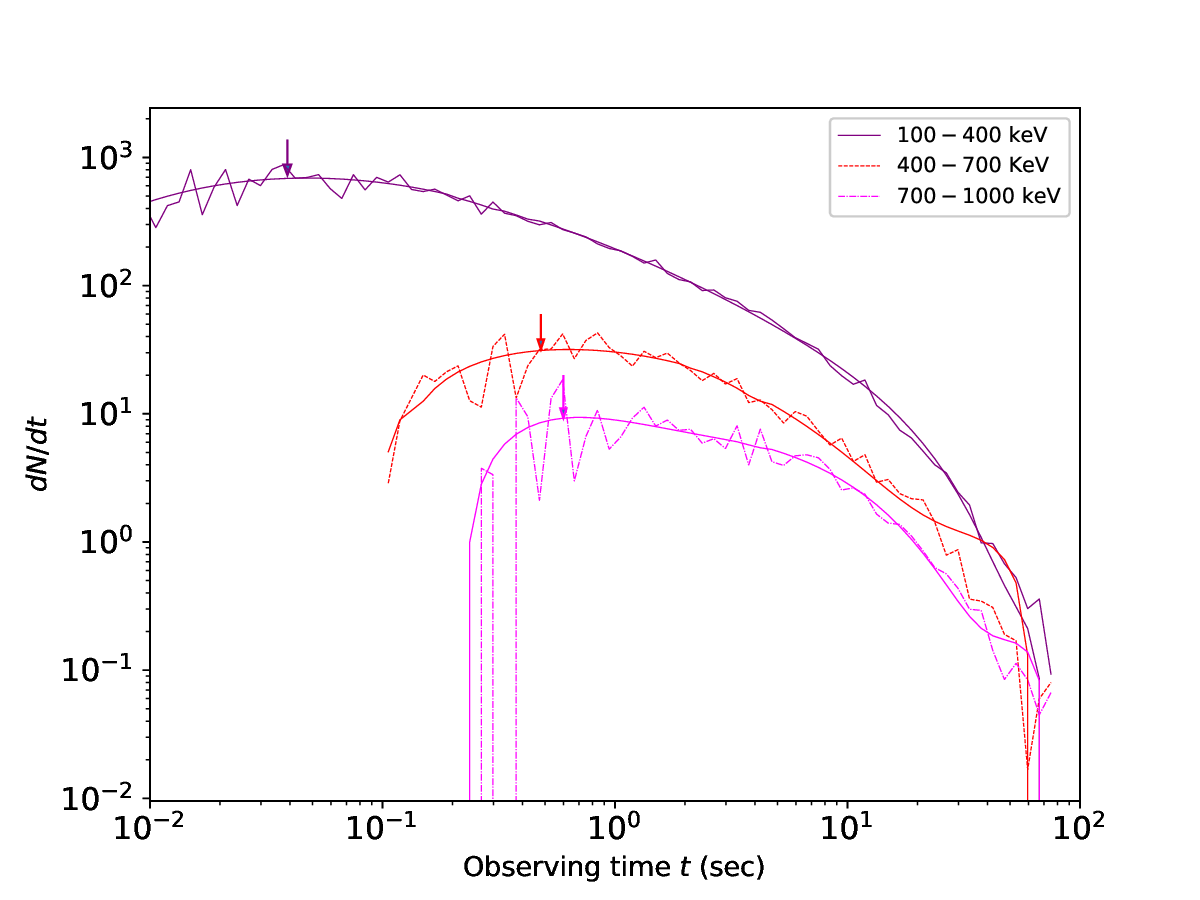}
\caption{Light curves for different energy channels showing positive lag in low energy range (Upper panel) and negative lag in high energy band (Lower panel). The light curves are over-plotted with Savitzky-Golay filter (solid lines). The vertical arrows represent the location of the peak of the light curves. The parameters are the same as in Figure \ref{lab_Photon_counts_g0_30_BB}.
}
\label{lab_Light_curve_positive_lag_BB_input}
 \end{center}
\end{figure}


\begin {figure}
\begin{center}
 \includegraphics[width=9cm, angle=0]{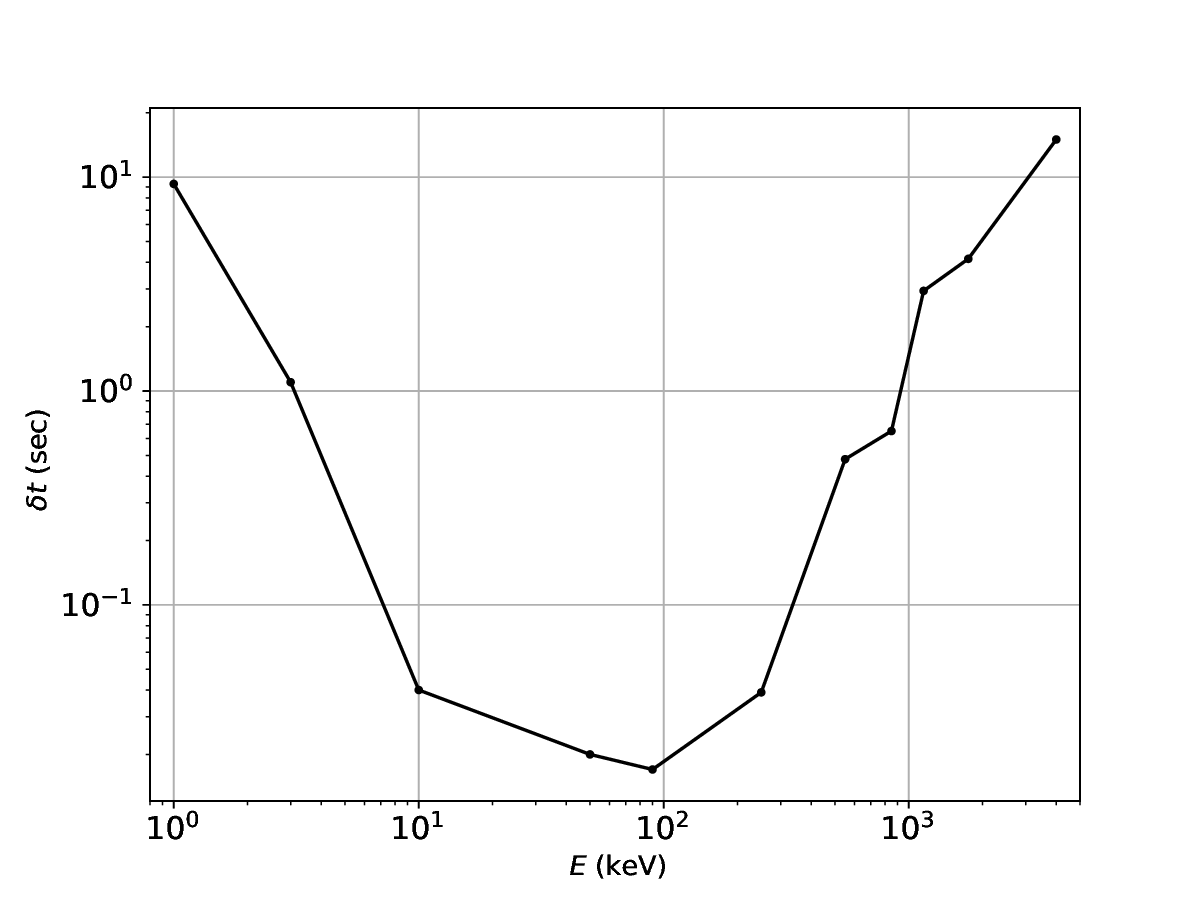}
\caption{Variation of spectral lag between different energy channels for $\Gamma_0=30$.
Other parameters are the same as in Figure \ref{lab_Light_curve_positive_lag_BB_input}. 
}
\label{lab_BB_lag}
 \end{center}
\end{figure}
Next, we consider a more realistic scenario, in which the jetted plasma is not cold but has a finite temperature that evolves along the jet as described in section \ref{sec_jet_stricture}. For the temperature calculation, we assume jet luminosity $L=10^{50}$~erg~s$^{-1}$. The inner jet angle is $\tj=0.01$ rad with $\Gamma_0=30$ and the jet shear parameter is $p=4$. The photons are assumed to be emitted at radius $r_{\rm inj} = \Rph/20$ with a thermal distribution determined by local fluid temperature $T'(r_{\rm inj}, \theta)$. The corresponding photon map along the $E-t$ plane is shown in Figure \ref{lab_Photon_counts_g0_30_BB} for an observer situated at $\to=\tj$. The first photon is received at $t\rightarrow 0$ with energy $E_{\rm Tr} = 10^2$ keV. For $E>E_{\rm Tr}$, photons arrive later as energies increase producing a negative lag while at energies $E<E_{\rm Tr}$, lower energy photons are delayed showing a positive spectral lag. 

Spectral lags are quantified by comparing the corresponding light curves in different energy channels. 
For the same parameters as in Figure \ref{lab_Photon_counts_g0_30_BB}, the simulated light curves for different energy channels, as seen by an observer situated along $\to=\tj$ are shown in Figure \ref{lab_Light_curve_positive_lag_BB_input}. The light curves show a fast rise followed by a power law decay, $dN_{ph}/dt\propto t^{-1}$ at later times in all bands. The spectral lag is quantified by the respective time lag seen between the peaks of the light curves at different energy channels.
The upper panel shows a positive lag at energies smaller than $100$ keV. The peak of the light curves denoted by black dotted curves shifts to earlier times for increasing energies $0-20$ keV (green solid), $20-40$ keV (black dashed) and $40-60$ keV (blue dashed-dotted).
For energies above $100$ keV, negative spectral lag is perceptible in the lower panel where the peak in the light curves shifts to later times as energies increase, $100-400$ keV (purple solid), $400-700$ keV (red dashed) and $700-1000$ keV (magenta dashed-dotted). The peaks associated with various light curves are marked with downward pointing arrows. They are estimated using Savitzky-Golay filter (over-plotted solid lines).

The corresponding time delay between the energy channels is plotted in Figure \ref{lab_BB_lag} for the parameters in Figure \ref{lab_Light_curve_positive_lag_BB_input}. The figure shows that the lag at lower energies is positive while it is negative at high energies. The negative lag is characterized by the positive slope while the positive lag is shown by the negative slope in the curve.
The transition between the two lags is at $E=E_{\rm 0}=100$ keV. For the parameters chosen, both the positive and negative lags reach the order of a few seconds at low (few keV) and high energies (few MeV). 

Note that $E_0$ is the energy at which the peak of the associated light curve has a minimum among all energy channels, while $E_{\rm Tr}$ is the energy of the first observed photon. These two energies are similar for an observed located at $\theta \leq \tj$: from Figure \ref{lab_Photon_counts_g0_30_BB}, $E_{\rm Tr}\sim$ 100~keV while from Figure $\ref{lab_BB_lag}$, $E_0\approx 100$~keV. 
$E_{\rm Tr}$ is the energy at which the burst is initially triggered and is a directly observable quantity. An analytic form of this transition energy was calculated\ in Equation \ref{eq_Etr}, $E_{\rm Tr} \approx 2 \Gamma_0 g_{a0}E_i'$. 
To proceed, we note that 
$E_i'$ directly corresponds to the gas temperature $T'$ (section \ref{sec_jet_stricture}). Thus the dependence of $E_{\rm Tr}$ on the jet parameters is obtained as,
\be 
E_{\rm Tr} \propto L^{-5/12} \Gamma_0^{8/3}.
\label{eq_E_0}
\ee 
The value of $E_{\rm Tr}$ thus depends on the jet Lorentz factor and the burst's isotropic luminosity. Hence when the jet luminosity is known, this equation hands a unique tool to determine the jet Lorentz factors or compare the Lorentz factors between various bursts. 


\section{Conclusions}
\label{sec_conclusions}
In this letter, we have addressed the problem of spectral lags observed during the GRB prompt phase. Positive spectral lag where the soft photons lag behind the hard photons is believed to arise from the geometrical curvature effect. However, the negative lag, so far, was lacking a clear physical understanding. 

Here we show that photon propagation within a structured jet provides a natural mechanism for producing such negative lags. Following our previous work on photon energy gain in an optically thick shearing plasma inside the jet \citep{2023ApJ...943L...3V}, the time the photons spend in gaining energy within the shear layers as well as their escape from higher altitudes altogether lead to their delay in the arrival time compared to the soft photons which forms the negative spectral lag.  For the negligible shear in the plasma along the jet axis, the conventional picture of a positive lag reappears where soft photons are delayed over the hard photons. 

Thus, combining both scenarios, the model gives a unified picture of the origin of positive and negative spectral lags in the GRB prompt phase. We further showed that in a typical GRB jet with shear, both lags can appear getting a flip from positive to negative lag at the transition energy $E_{\rm Tr}$. This feature is consistent with the GRBs that show such lag transitions in their prompt phase observations. The spectral lag transition from positive to negative is observed in a fraction of GRBs  \citep{2017ApJ...834L..13W, 2022PASA...39....1G,2023ApJ...942...67L, 2022MNRAS.511.1694G}. 
Out of 135 long GRBs with known redshifts observed by Fermi-GBM, around a quarter of them report this transition \citep{2022ApJ...935...79L}. The reported transition energy is found in the range 100 keV - 1 MeV, while the values of energies in the negative spectral lag range between a hundred keV to around 10 MeV. The magnitudes of the spectral lags are reported from a fraction of a second to a few seconds. These are in excellent agreement with the range of values we have obtained in this letter. Thus the structured jets provide a viable physical scenario to understand the delay of high-energy photons.

Other than explaining the lags, the model provides a unique tool to quantify the parameters associated with the bursts.
The transition energy at which the positive lag switches to negative lag prominently depends upon the Lorentz factor $\Gamma_0$ as well as the luminosity $L_0$ (Equation \ref{eq_E_0}), allowing one to constrain the jet Lorentz factor from direct observations of the spectral lags.
The range of energies during negative delay depends upon the assumed shear strength. For values $p\le 2$, no shear is obtained and only positive lag should persist in such GRB jets. For significant shear, the magnitude of the delay 
is largely independent of shear parameter $p$.
We will continue this study, constraining the physical characteristics of different bursts as applications of our model.

The observed lag time may provide an independent clue on the jet Lorentz factor.
The obtained duration of the spectral lags in this study, i.e., a few seconds, match the observations for an inner jet Lorentz factors of around 30 (see Figure \ref{lab_BB_lag}). This result therefore suggests that a very high Lorentz factors ($\Gamma_0\ge 100$) may not be likely in many long GRBs, consistent with the findings of \cite{2022NatCo..13.5611D}. 
As recently reported in the Fermi-LAT observations, photons at even higher energies (100 MeV - 10 GeV) generally arrive at a delay compared to the soft photons in the vast majority of GRBs \citep{2013ApJS..209...11A}. Thus we argue that the transition from positive lag to negative lag should be a universal feature in the GRB prompt phase, however, the energy range obtained by us quantitatively covers only the results of Fermi - GBM ($\le 10 $ MeV). 





\section*{Acknowledgments}
We acknowledge the support from the European Union (EU) via ERC consolidator grant 773062 (O.M.J.) and to the Israel Space Agency via grant \#6766. S.I. is supported by DST INSPIRE Faculty Scheme (IFA19-PH245), SERB SRG Grant (SRG/2022/000211) and Centre of High Performance Computing, IISER Thiruvananthapuram, Kerala, India

\appendix
\section{Appearance of photosphere for various observers}
\begin {figure}
\begin{center}
  \includegraphics[width=6cm, angle=0]{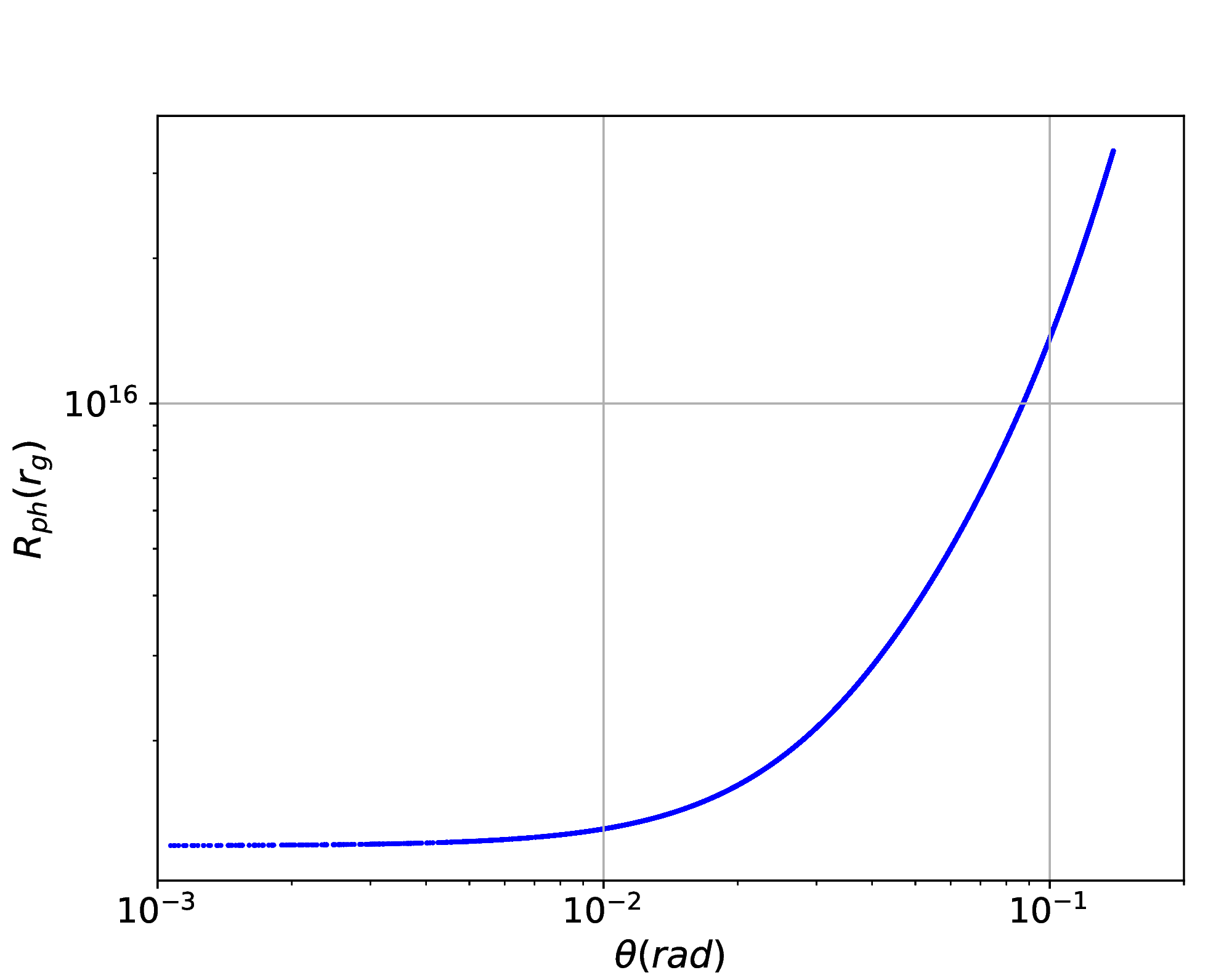} 
  \includegraphics[width=6cm, angle=0]{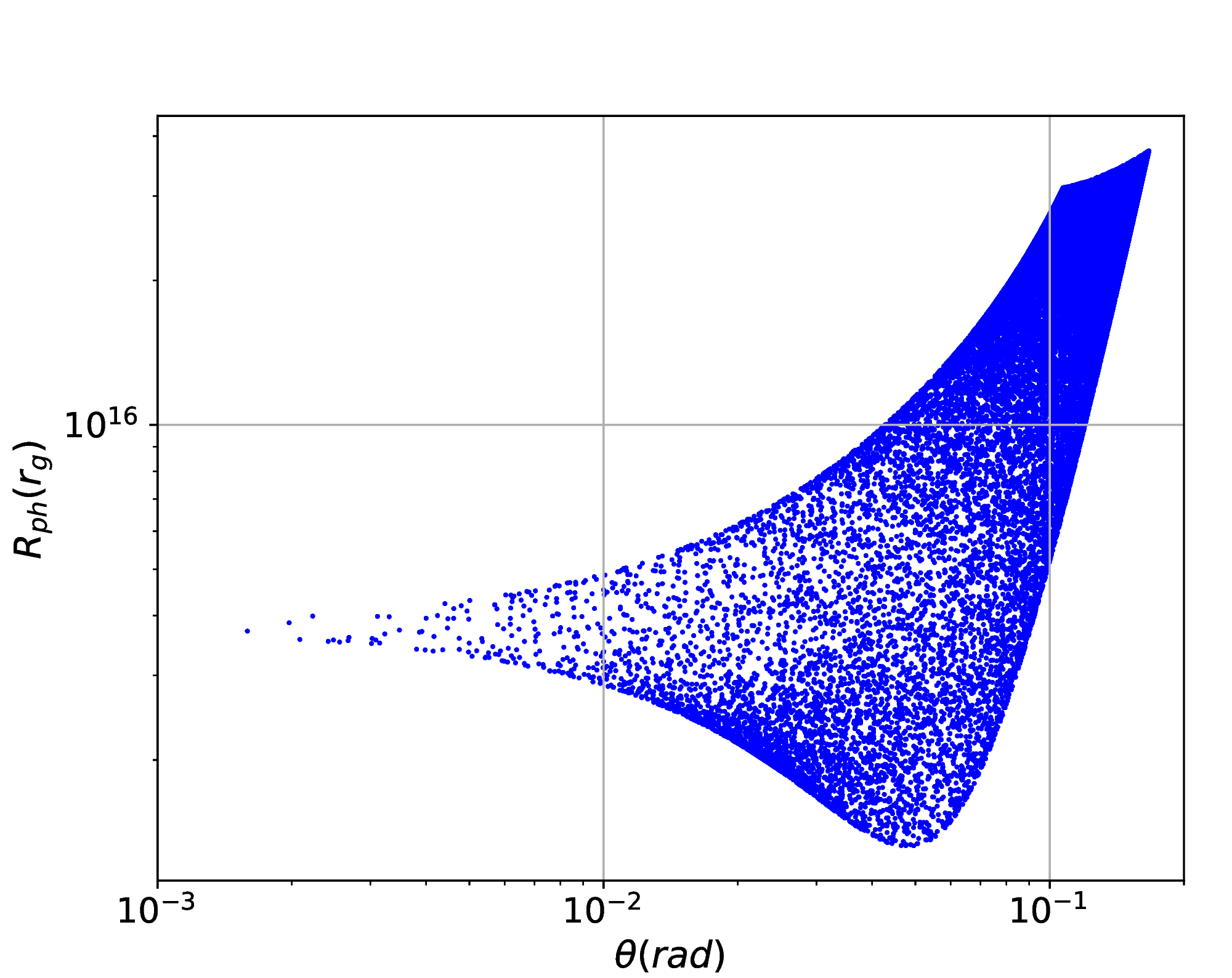} 
   \includegraphics[width=6cm, angle=0]{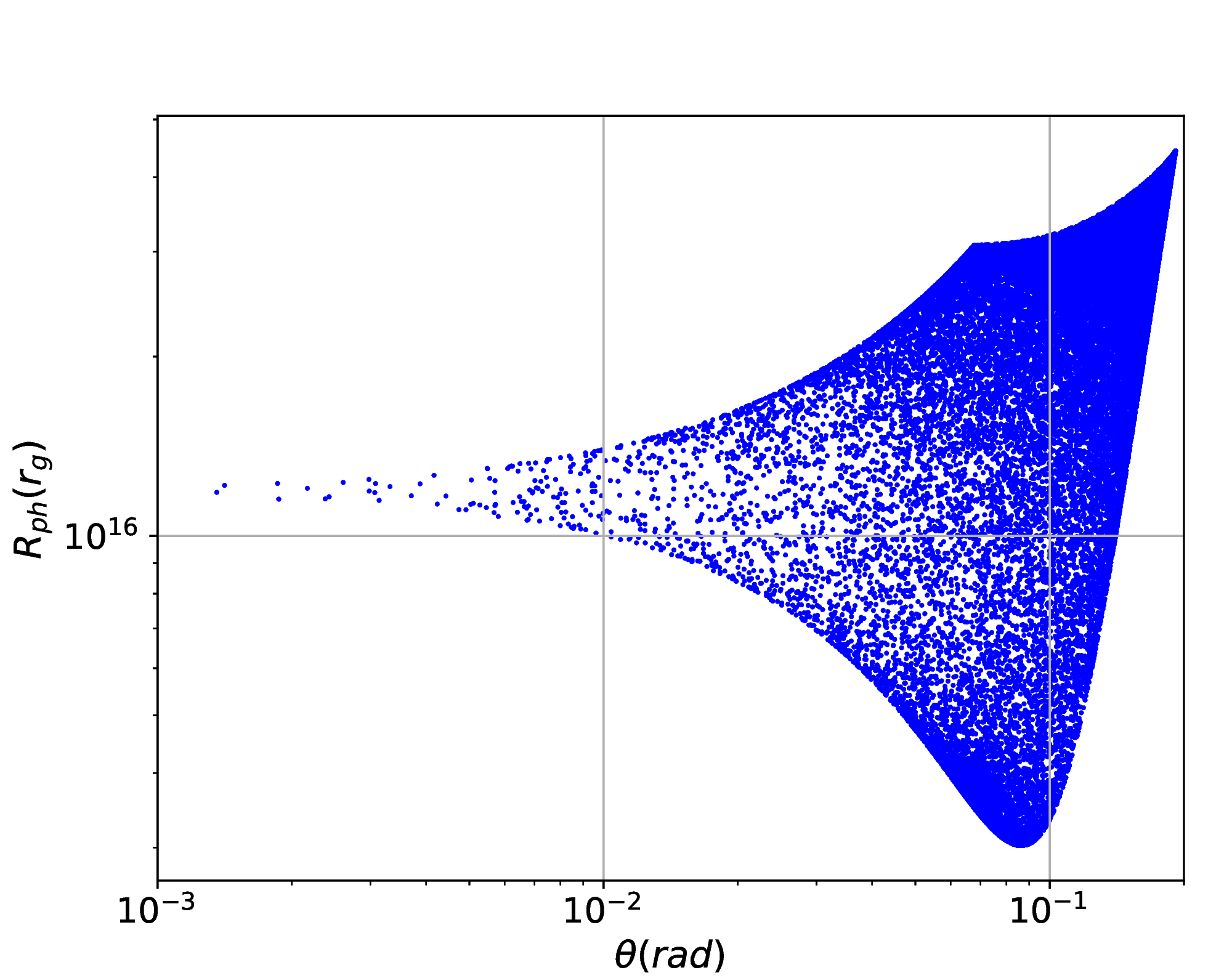} 
\caption{Photospheric radius for viewing angles $\to=0,0.05$ and $0.10$ with $p=2.0, \tj=0.10$ rad (Top to bottom), $\Gamma_0=50$ and $L=10^{52}$~erg~s$^{-1}$.
}
\label{lab_tp0}
 \end{center}
\end{figure}

The jet system possesses an azimuthal symmetry with parameters being independent of $\phi$. However, the photosphere seen by an off-axis observer ($\to>0$) is not axisymmetric. As described in \cite{2024ApJ...972...40V}, the optical depth for scattering is  

\be 
\tau = \int_{0}^{\tl} \frac{n_e' \Gamma r \sigma_T [1-\beta \cos \tilde{\tl}]}{2\sin \tilde{\tl}}d\tilde{\tl}.
\label{eq_phot}
\ee
Here, $\tl$ is the angle between the local fluid direction and the observer's angular location (see lower panel of Figure \ref{lab_geom1}). For fluid element at the jet axis, $\theta=0$, $\tl=\to$. for $\to>0$, the angle $\tl$ depends on the local fluid direction and hence it is a function of both $\theta$ and $\phi$.
The surface of the photosphere $r=\Rph$ is, by definition, the surface from which the optical depth $\tau=1$, and is calculated using the above equation. Hence for given jet parameters and an observer's location $\to$, a unique surface of the photosphere is obtained. The comoving electron density is $n_{\rm e}'=L/4\pi m_{\rm p} c^3 \beta \Gamma^2 r^2$, thus for $\tau=1$, $r=\Rph$ can be expressed as,

\be 
\Rph =  \frac{L\sigma_T}{8\pi m_pc^3}\int_{0}^{\tl}\frac{[1-\beta \cos \tilde{\tl}]}{\beta \Gamma r \sin \tilde{\tl}}d\tilde{\tl}
\ee 
The obtained photosphere is shown in Figure \ref{lab_tp0} for parameters $\Gamma_0=50$, $\tj=0.1$ rad, $p=2.0$ and $L=10^{52}$ ~erg~s$^{-1}$ for observer's angular locations being angles $\to=0,0.05$ and $0.10$ rad (upper left, upper right and bottom panel respectively), and all other parameters are the same as in Figure \ref{lab_Photon_counts_g0_100_tj_0.01_L_1e52}. For an on-axis observer, the distribution is a single curve with $\Rph$ monotonously increasing with polar angle $\theta$, while for an off-axis observer, there is a range of radii for each angle, $\theta$. This is due to the different values of $\phi$, that introduce a family of curves. The lowermost curve shows the photosphere for $\phi=\phi_{\rm o} = 0$ while the top surface corresponds to $\phi=\pi$. The minimum of the photosphere for $\phi=0$ follows a variation of $\theta=\to$ while it is at $\theta=0$ for an on-axis observer.

\bibliography{ref1}{}
\bibliographystyle{aasjournal}
\end{document}